\documentclass{article}

\begin{document}

\title{Analysis of a mathematical model for interactions between T cells and 
macrophages}

\author{Alan D. Rendall\\
Max-Planck-Institut f\"ur Gravitationsphysik\\
Albert-Einstein-Institut\\
Am M\"uhlenberg 1\\
14476 Potsdam, Germany}

\date{}

\maketitle

\begin{abstract}
The aim of this paper is to carry out a mathematical analysis of a system 
of ordinary differential equations introduced by R. Lev Bar-Or to model the 
interactions between T cells and macrophages. Under certain restrictions
on the parameters of the model, theorems are proved about the number of 
stationary solutions and their stability. In some cases the existence of
periodic solutions or heteroclinic cycles is ruled out. Evidence is 
presented that the same biological phenomena could be equally well 
described by a simpler model.
\end{abstract}

\section{Introduction}

Autoimmune diseases result in a great deal of suffering for affected 
individuals and huge costs for society. Notable examples are multiple
sclerosis, rheumatoid arthritis and type I diabetes. In these diseases the 
ability of the immune system to distinguish between self and non-self is 
compromised, with the result that host tissues are attacked and damaged.
It is important to get a better understanding of the processes involved and 
one way to do so is to introduce theoretical models of the immune system,
in particular mathematical models.

There has been a lot of work on mathematical modelling of interactions
of the immune system with pathogens. See for instance the book of Nowak
and May \cite{nowak} which concentrates on the case of HIV. Autoimmune 
diseases do not need to involve any pathogens, although pathogens might 
contribute to them indirectly. For this reason it would be interesting to 
have models for the intrinsic workings of the immune system where non-self 
antigens play no direct role. Apparently few models of this type exist in 
the literature. One example is introduced
in a paper of Lev Bar-Or \cite{levbaror}. It is a system of four ordinary
differential equations which describes the interactions of T cells and 
macrophages by means of the cytokines they produce. The aim of this paper is 
to investigate what can be said about the properties of the solutions of 
this system on the level of mathematical proofs. 

T cells are white blood cells which mature in the thymus. One type of
T cell, the T helper cell, helps to direct the activity of other immune 
cells. These cells are also known as CD4${}^+$ since they carry the 
surface molecule CD4. T cells communicate with other cells by secreting 
soluble substances known as cytokines. Another important type of blood cell 
is the macrophage which ingests pathogens and cell debris through 
phagocytosis. Macrophages also secrete cytokines. Both T cells and 
macrophages react in various ways to the cytokines which are present in their 
surroundings. The list of known cytokines is long and each of them has its 
own characteristics in terms of which types of cells secrete it and what 
effects it has on cells which detect its presence. An idea of the complexity 
of this signalling system can be obtained from \cite{levbaror}.

It is common to distinguish between two types of T helper cells, known as
Th1 and Th2, according to the cytokines they produce. There may be overlaps
but roughly speaking it may be supposed that there is one set of cytokines
which are called type 1 and are produced by Th1 cells and another called 
type 2 which are produced by Th2 cells. Macrophages produce cytokines of 
both types. The basic quantities in the equations of \cite{levbaror} are
average concentrations corresponding to type 1 and type 2 cytokines produced 
by T cells and macrophages. These four concentrations are denoted by $C^T_1$, 
$C^T_2$, $C^M_1$ and $C^M_2$. The populations of different cell types do not 
occur directly in the system. The quantities which have a chance of being 
measured directly are $C_1=\frac12(C^T_1+C_1^M)$ and $C_2=\frac12(C^T_2+C_2^M)$. 

There are cases where the immune response is dominated by either Th1 or Th2 
cells. This may be important in order to effectively combat a particular
pathogen. For instance a sufficiently strong Th1 response is necessary for
containing or eliminating a tuberculosis infection \cite{wigginton}, 
\cite{marino}. An inappropriate balance between these two states can also 
contribute to autoimmune disorders. It has, for instance, been suggested that 
multiple sclerosis is associated with an immune response which is biased 
towards Th1. See \cite{lassmann} for a critical review of this idea. In the 
context of the model it is said that there is Th1 or Th2 dominance if 
$C_1>C_2$ or $C_2>C_1$, respectively.

Another important function of macrophages which plays a role in the model
of \cite{levbaror} is the presentation of antigens. Small peptides
which result from the digestion of material taken up by a macrophage
are presented on its surface in conjunction with MHC II molecules. (Major
histocompatibility complex of class II.) This can stimulate T cells
which come into contact with the macrophage.

It has not been possible to give a complete analysis of
the dynamics of the system of \cite{levbaror}.  A certain inequality on
the parameters of the system gives rise to a regime where there
is a unique stationary solution which acts as an attractor for all solutions
as $t\to\infty$. This is proved in Theorem 1. For a more restricted set of
parameters it is possible to show (Theorem 2) that each solution converges
to a stationary solution which in general depends on the solution considered.
With further restrictions on the parameters it is shown that there are 
between one and three stationary solutions and the subsets of parameters
for which different numbers of stationary solutions occur are described.
This is the content of Theorem 3. In Theorems 2 and 3 the coefficients 
describing antigen presentation are set to zero. One situation 
where information can be obtained on the dynamics including antigen 
presentation is analysed in Theorem 4. 

The analysis which has been done has uncovered no evidence that the inclusion
of the effect of macrophages makes an essential difference to the behaviour of
solutions. The types of qualitative behaviour which have been proved to occur
in this paper and those which are shown in the figures in \cite{levbaror} can
be found in a truncated system which only includes the effect of T cells. 
This is discussed in section \ref{Tcellonly}.

\section{Analysis of the dynamical system}

In what follows it will be convenient to use a notation which is 
more concise than that of \cite{levbaror}. Let $x_1$, $x_2$, $x_3$, $x_4$, 
$z_1$ and $z_2$  denote $C^T_1$, $C^T_2$, $C^M_1$, $C^M_2$, $x_1+x_3$ and 
$x_2+x_4$ respectively. The basic dynamical system is:
\begin{equation}\label{basic}
\frac{dx_i}{dt}=-d_ix_i+g(h_i);\ \ \ \ i=1,2,3,4.
\end{equation}
The $d_i$ are positive constants. The function $g$ is given 
by 
\begin{equation}
g(x)=\frac12 (1+\tanh (x-\theta))
\end{equation}
where $\theta$ is a constant. It 
satisfies the relations $g(x+\theta)+g(-x+\theta)=1$ and 
$g'(x)=2g(x)(1-g(x))$. Hence $g(\theta)=\frac12$ and $g'(\theta)=\frac12$.
The functions $h_i$ are defined by 
$h_i=\sum_{j}a_{ij}x_j$ for some constants
$a_{ij}$. The coefficients in (\ref{basic}) satisfy the following conditions
\begin{enumerate}
\item{$a_{ij}=b_{ij}+c_{ij}$, $i=1,2$, for some coefficients 
$b_{ij}$ and $c_{ij}$}
\item{$b_{1j}=-b_{2j}$ for all $j$.}
\item{$b_{11}>0$, $b_{13}>0$, $b_{12}<0$ and $b_{14}<0$}
\item{The ratio $c_{2j}/c_{1j}$ is independent of $j$ and positive.}
\item{$c_{11}>0$ and $c_{13}>0$}
\item{$a_{3j}=-a_{4j}$ for all $j$}
\item{$a_{31}>0$, $a_{33}>0$, $a_{32}<0$, $a_{34}<0$}
\end{enumerate}  
The sign conditions encode the fact that the effect of type 1 cytokines 
on cells is to increase their production of type 1 cytokines and to decrease
their production of type 2 cytokines, while the effect of type 2 cytokines is
exactly the opposite. The coefficients $c_{ij}$ encode the effects of antigen
presentation. No assumption is made on the signs of $c_{12}$ and $c_{14}$. 
There are eighteen parameters in the model which are only constrained by some 
positivity conditions. The quantities $z_1$ and $z_2$ represent total 
concentrations of type 1 and type 2 cytokines. The biologically relevant 
region $\cal B$ is that where all the $x_i$ are non-negative. The function 
$g$ is strictly positive. Hence if one of the variables $x_i$ vanishes at some 
time its derivative at that time is strictly positive. It follows that 
$\cal B$ is positively invariant under the evolution defined by the system. 
If some $x_i$ is greater than or equal to $d^{-1}_i$ on some time interval 
then $x_i$ is decreasing at a uniform rate during that time. It follows that 
all solutions exist globally to the future and enter the region ${\cal B}_1$ 
defined by the inequalities $x_i\le d^{-1}_i$ after finite time. Thus in order 
to study the late-time behaviour of any solution starting in $\cal B$ it is 
enough to consider solutions starting in ${\cal B}_1$.  In fact any solution 
enters the interior of ${\cal B}_1$ after finite time. 

\noindent
{\bf Theorem 1} For any value of the parameters the system (\ref{basic}) has
at least one stationary solution. If
\begin{equation}\label{contractive}
\sup_i\sum_j |a_{ij}|< 2\inf_i d_i.
\end{equation}
then there is only one stationary solution and all solutions converge to
it as $t\to\infty$. 

\noindent
{\bf Proof} That the system always has at least one stationary solution 
follows from the Brouwer fixed point theorem, (cf. \cite{hale}, Theorem I.8.2).
A stationary solution of the system satisfies
\begin{equation}
x_i=d_i^{-1}g(h_i).
\end{equation} 
To prove the second part of the theorem consider the following estimate
\begin{equation}
|d_i^{-1}g(h_i(y))-d_i^{-1}g(h_i(x))|\le (2d_i)^{-1}|h_i(y)-h_i(x)|\le 
(2d_i)^{-1}\sum_j |a_{ij}||y_j-x_j|.
\end{equation}
The first of these inequalities uses the mean value theorem and the fact 
the derivative of $g$ is nowhere greater than one half. If (\ref{contractive})
holds then the mapping $\{x_i\}\mapsto \{d_i^{-1}g(h_i)\}$ maps ${\cal B}_1$ to 
itself and is a contraction in the 
maximum norm. Hence it has a unique fixed point. It follows that when the 
coefficients of the system satisfy the restriction (\ref{contractive}) the 
system has exactly one stationary solution. If $x(t)$ and $y(t)$ are two
solutions then under the assumption (\ref{contractive}) it can be shown that 
$|x-y|$ decays exponentially as $t\to\infty$. Thus all solutions converge to 
the unique stationary solution as $t\to\infty$. A related statement is that 
if $x_*$ is the unique stationary solution then $|x-x_*|^2$ is a Lyapunov 
function. 

\vskip 10pt
A limiting case of (\ref{basic}) is that where all $c_{ij}$ vanish. This will 
be called the zero MHC system. The pattern of signs in the coefficients in the 
zero MHC system is such that the change of variables $\tilde x_i=(-1)^{i+1}x_i$
leads to a system $\frac{d\tilde x_i}{dt}=f(\tilde x_j)$ satisfying 
$\frac{\partial f_i}{\partial \tilde x_j}> 0$ for all $i\ne j$. This means 
that the  dynamical system for the $\tilde x_i$ is cooperative \cite{hirsch}. 
This pattern
of signs corresponds to what is referred to as a \lq community with limited 
competition\rq\ or \lq competing subcommunities of mutualists\rq\ in 
\cite{smith86}. It implies, using the Perron-Frobenius theorem, that the
linearization at any point of the vector field defining (\ref{basic}) has
a real eigenvalue of multiplicity one which is greater than the modulus of any
other eigenvalue. Notice that in the special case of the zero MHC system
where all $d_i$ are equal to $d$ the linearization has two eigenvalues equal 
to $-d$. As a consequence all the eigenvalues of the linearization must be
real in this case. It is not clear how these facts about the eigenvalue 
structure can be used to help to understand the dynamics. The condition on the 
coefficients $d_i$, which says that the rate of degradation of 
different cytokines is exactly equal, would not be true with biologically 
motivated parameters. Nevertheless it is not unreasonable to assume that these 
coefficients are approximately equal and that the model with equal 
coefficients is not a bad approximation to the situation to be modelled. An 
assumption of this type is made in the model of \cite{marino}.

Call the system obtained by assuming $c_{ij}=0$, $d_1=d_2$, $d_3=d_4$ and 
$\theta=0$ in (\ref{basic}) System 2 while (\ref{basic}) itself is System 1. 
The condition $\theta=0$ implies that $g(x)+g(-x)=1$. Adding the equations for 
$i=1$ and $i=2$ then gives
\begin{equation}
\frac{d}{dt}(x_1+x_2)=-d_1(x_1+x_2)+1.
\end{equation}
Similarly
\begin{equation}
\frac{d}{dt}(x_3+x_4)=-d_3(x_3+x_4)+1.
\end{equation}
These equations can easily be analysed. An invariant manifold $S_1$ is defined 
by  $x_1+x_2=d_1^{-1}$ and $x_3+x_4=d_3^{-1}$. Substituting this back into the
full system gives a two-dimensional system written out below which will 
be called System 3. The $\omega$-limit set of any solution of System 2 is
contained in the invariant manifold $S_1$. Passing from System 1 to System 3 
means setting some parameters to zero and then restricting to a 
two-dimensional invariant manifold of the resulting system. System 3 may not 
include all the most interesting dynamics exhibited by solutions of the system 
of \cite{levbaror} but can be perturbed to get information about cases where 
the effect of the MHC is small but non-zero. The phase portrait of System 2 
follows immediately from that of System 3. The explicit form of System 3 is:
\begin{eqnarray}\label{system3}
&&\frac{dx_1}{dt}=-d_1 x_1+g((a_{11}-a_{12})x_1
+(a_{13}-a_{14})x_3+a_{12}d_1^{-1}+a_{14}d_3^{-1})\\
&&\frac{dx_3}{dt}=-d_3 x_3+g((a_{31}-a_{32})x_1
+(a_{33}-a_{34})x_3+a_{32}d_1^{-1}+a_{34}d_3^{-1})
\end{eqnarray}
Note that the coefficients of $x_1$ and $x_3$ in these equations which are
linear combinations of the $a_{ij}$ are all positive. The constant terms are 
negative. This is a cooperative system. This fact together with the fact that 
the dimension of the system is two implies that each solution converges to 
a limit as $t\to\infty$ \cite{hirsch}. In other words the $\omega$-limit
set of each solution is a single point. Furthermore, there are no homoclinic 
orbits or heteroclinic cycles. What is not clear in general is how many 
stationary solutions there are. Some of the conclusions of the above 
discussion can be summarized as follows:

\noindent
{\bf Theorem 2} If $c_{ij}=0$ for all $i$, $j$, $d_1=d_2$, $d_3=d_4$ and 
$\theta=0$ then every solution of (\ref{basic}) converges to a stationary 
solution satisfying $x_1+x_2=d_1^{-1}$ and $x_3+x_4=d_3^{-1}$ as $t\to\infty$. 
There are no homoclinic orbits or heteroclinic cycles.

Consider now the special case of System 3 obtained by setting $a_{1j}=a_{3j}$ 
for all $j$ and $d_1=d_3$. Call it System 4. This corresponds to the 
assumptions that the T cells and macrophages have identical properties with 
respect to their death rate and their interactions with cytokines. This 
system consists of two copies of a single equation. The quantity $x_1-x_3$
decays exponentially. The further assumption $x_1=x_3$ on the initial data, 
which means that there are equal numbers of T cells and macrophages, gives 
rise to a single ODE, call it the toy model. It is the same equation which
occurs twice in System 4. It is of the form
\begin{equation}\label{pretoy}
\frac{dx}{dt}=-d_1x+g(ax-b)
\end{equation} 
where $a=a_{11}-a_{12}+a_{13}-a_{14}$ and $b=-(a_{12}+a_{14})d_1^{-1}$. Note
that $a$ and $b$ are positive. Here the zero MHC condition has been used.
This can be simplified by defining $x'=d_1 x$, $t'=d_1 t$ and 
$a'=\frac{a}{d_1}$. Suppressing the primes leads to
\begin{equation}\label{toy}
\frac{dx}{dt}=-x+g(ax-b)
\end{equation}
Denote the right hand side of (\ref{toy}) by $h(x)$, suppressing the parameter 
dependence. Stationary solutions of the toy model are given by solutions of 
the equation
\begin{equation}\label{toystat}
g(ax-b)=x.
\end{equation}
Since $0<g<1$ all solutions of this equation are contained in the interval
$(0,1)$. The right hand side of this equation takes the value zero at $x=0$ 
and the value one at $x=1$. Thus by the intermediate value theorem 
(\ref{toystat}) has at least one solution on the interval of interest. 
Now
\begin{equation}\label{hump}
\frac{d}{dx}(g(ax-b)-x)=2ag(ax-b)[1-g(ax-b)]-1.
\end{equation}
The first term on the right hand side of (\ref{hump}) is symmetric about
$x=\frac{b}{a}$ where it attains its maximum value $\frac{a}{2}$. Thus 
this derivative vanishes at zero, one or two points of the real line according 
to whether $a<2$, $a=2$ or $a>2$. For $a>2$ call the zeroes of the derivative
$x_1$ and $x_2$ with $x_1<x_2$. It can be concluded that for any value of $a$
there are at most three solutions of (\ref{toystat}). All these solutions 
must be within the interval of interest. If there are exactly two solutions 
then one of them must correspond to a point where the right hand side of 
(\ref{toy}) and its derivative vanish simultaneously. For otherwise the value 
of this function at one would be negative for large $x$, a contradiction. For
a given value of $a>2$ there is precisely one value of $b$ for which 
the equation $-x_1+g(ax_1-b)=0$ holds. Call it $b_1(a)$. Define $b_2(a)$ 
similarly, replacing $x_1$ by $x_2$.
Then $b_1(a)$ and $b_2(a)$ are the only values of $b$ for which there is a 
simultaneous solution of $g(ax-b)=x$ and $ag'(ax-b)=1$. For all $a>2$ the 
inequality $b_1(a)<b_2(a)$ holds. As $a\to 2$ both tend to one. The 
union of the graphs of $b_1$ and $b_2$ divides the $(a,b)$-plane into two
regions. On each of these the number of stationary solutions is constant 
and equal to one or three. On the region including points with $a<2$ it is
obviously one. Considering points where $b=\frac12$ and $a$ slightly larger 
than two makes makes it clear that it is three on the other region. 
Information can also be obtained by simple geometric considerations on 
the position of the stationary points. If $a<2$ and $b<\frac{a}2$ then 
the stationary point has $x<\frac12$ while for $b>\frac{a}2$ it satisfies
$x>\frac12$. If $a>2$ and there are three stationary points then the 
central (unstable) one satisfies $x<\frac12$ for $b<\frac{a}2$ and 
$x>\frac12$ for $b>\frac{a}2$.  

When there is only one solution it is a hyperbolic sink. When there are three
solutions the two outermost are hyperbolic sinks while the intermediate one
is a hyperbolic source. If the parameter $b$ is held fixed at some value 
less than one and $a$ is varied then at the value of $a$ where $b_1(a)=b$ 
there is a fold bifurcation. To see this note that
\begin{eqnarray}
&&\frac{\partial h}{\partial x}=-1+ag'(ax-b)   \\
&&\frac{\partial h}{\partial a}= xg'(ax-b)=2xg(ax-b)(1-g(ax-b))   \\ 
&&\frac{\partial^2 h}{\partial x^2}=a^2g''(ax-b)
=4a^2g'(ax-b)(1-2g(ax-b))
\end{eqnarray}
From the first equation it follows that at a critical point $x$ the quanitity
$ag'(ax-b)$ must be equal to one. Hence the derivative with respect to $a$
does not vanish there. The second derivative with respect to $x$ can only 
vanish there if $g(ax-b)=\frac12$, which implies that $x=\frac{b}{a}$,
$x=\frac12$ and $b=1$. The situation for $b>1$ is similar to that for $b<1$. 
To understand the case $b=1$ note that there is a cusp bifurcation in $(a,b)$ 
at the point $(2,1)$. This can be shown using the facts that
\begin{eqnarray}
\frac{\partial h}{\partial b}=&&-g'(ax-b),         \\
\frac{\partial^2 h}{\partial x\partial b}=&&-ag''(ax-b),  \\
\frac{\partial^2 h}{\partial x\partial a}=&&axg''(ax-b),
+g'(ax-b)\nonumber \\
\frac{\partial^3 h}{\partial x^3}=&&g'''(ax-b) \\
=&&4a^3g''(ax-b)(1-2g(ax-b))-8a^3(g'(ax-b))^2.
\end{eqnarray}
Hence $\frac{\partial^3 h}{\partial x^3}$ and 
$\frac{\partial h}{\partial a}\frac{\partial^2 h}{\partial x\partial b}-
\frac{\partial h}{\partial b}\frac{\partial^2 h}{\partial x\partial a}$
are both non-vanishing and Theorem 8.1 of \cite{kuznetsov} applies.

If we remove the condition $x_1=x_3$ then $x_1-x_3$ decays exponentially. The 
qualitative nature of the dynamics of System 4 with the given restrictions on 
the parameters is then clear. This in turn gives information about the 
phase portrait of System 2 with these values of the parameters. Converting 
back to the original variables has the effect of replacing $a$ by $a/d_i$ in 
these criteria. Some of these results are summarized in the following theorem:

\noindent
{\bf Theorem 3} If in Theorem 2 it is additionally assumed that 
$a_{1j}=a_{3j}$ for all $j$ and $d_1=d_3$ then the number of stationary 
solutions of (\ref{basic}) is between one and three for any values of the 
parameters. It is one whenever $a_{11}-a_{12}+a_{13}-a_{14}\le 2$. There is a 
non-empty open set where it is three. The set where it is two is a union of 
two smoothly embedded curves which is the boundary between the sets where it 
is one and three. 

Consider a choice of parameter set for (\ref{basic}) for which there are
three hyperbolic stationary points. It follows by a simple stability argument 
that there is an open neighbourhood of this parameter set in parameter space 
such that for all parameters in this neighbourhood there are precisely three 
hyperbolic stationary points. If the original set of stationary points
consists of two sinks and one saddle then there is an open neighbourhood 
where this property persists.

In the context of this theorem the question of Th1 or Th2 dominance 
can be examined for stationary solutions. It is of interest to know whether 
changing the parameters in the system can cause a switch from one type of
dominance to the other. If this happens there must be some values of the 
parameters for which there is a stationary solution with $z_1=z_2$. 
Under the hypotheses of the Theorem 2 $x_2=d_1^{-1}-x_1$ and 
$x_4=d_1^{-1}-x_3$
for any stationary solution. Hence $z_2=2d_1^{-1}-z_1$. If $z_1=z_2$ then 
this implies that $z_1=d_1^{-1}$ and $x_1=\frac12 d_1^{-1}$. In terms of the 
new variable introduced in (\ref{toy}) this means that $x=\frac12$. The
above discussion gives information about Th1 or Th2 dominance for various
parameter values satifying the restrictions in the statement of Theorem 3. 
What is of most interest for the applications is the position of the stable 
stationary points. In particular it is clear that there exist parameter 
values where there are stationary solutions with both types of dominance. 

Now another set of simplified versions of the system will be studied. The 
phase portraits given in Fig. 2 in \cite{levbaror} relate to one special
case of this kind. In the notation used here it is defined by the following
restrictions on the parameters: $d_i=1$ for all $i$, $\theta=0$, all
coefficients $b_{ij}$ with $j=1$ or $j=3$ are equal, all coefficients $a_{ij}$ 
with $j=2$ or $j=4$ are equal, the coefficients $c_{ij}$ are equal for all $i$ 
and $j$. This reduces the number of free parameters to three. To simplify the 
notation let $A=b_{11}$, $B=-b_{12}$, $C=c_{11}$. These are all positive 
constants. For Fig. 2 in \cite{levbaror} two sets of values for the 
coefficients are considered. In both cases
$C=0.5$. In Fig. 2a $A=0.4$ and $B=0.5$ while in Fig. 2b $A=0.6$ and 
$B=0.65$. With these assumptions System 1 becomes:
\begin{eqnarray}\label{ABCsystem}
&&\frac{dx_1}{dt}=-x_1+g((A+C)(x_1+x_3)+(-B+C)(x_2+x_4))\label{abc1}\\
&&\frac{dx_2}{dt}=-x_2+g((-A+C)(x_1+x_3)+(B+C)(x_2+x_4))\label{abc2}\\
&&\frac{dx_3}{dt}=-x_3+g(A(x_1+x_3)-B(x_2+x_4))\label{abc3}\\
&&\frac{dx_4}{dt}=-x_4+g(-A(x_1+x_3)+B(x_2+x_4))\label{abc4}
\end{eqnarray}
Call this System 5. If $A+B+2C<1$ then there is at
most one stationary point of this system. This can be proved in the same 
way as Theorem 1.  Adding the equations in pairs 
leads to a closed system for the two experimentally accessible quantities 
$z_1$ and $z_2$. These are the variables which are plotted in Fig. 2 of 
\cite{levbaror}.
\begin{eqnarray}
&&\frac{dz_1}{dt}=-z_1+g((A+C)z_1+(-B+C)z_2)+g(Az_1-Bz_2)\label{z1evol}    \\
&&\frac{dz_2}{dt}=-z_2+g((-A+C)z_1+(B+C)z_2)+g(-Az_1+Bz_2)\label{z2evol}
\end{eqnarray}
Call this System 6. Denote the right hand sides of (\ref{z1evol}) and 
(\ref{z2evol}) by $f_1(z_1,z_2)$ and $f_2(z_1,z_2)$ respectively.
For this system uniqueness of the stationary solution 
follows from the assumption that the coefficients satisfy $A+B+C<1$. 
If in addition the coefficient $C$ is assumed to vanish then this reduces to 
\begin{eqnarray}
&&\frac{dz_1}{dt}=-z_1+2g(Az_1-Bz_2)\label{z1evol3}\\
&&\frac{dz_2}{dt}=-z_2+2g(-Az_1+Bz_2)\label{z1evol4}
\end{eqnarray}
Call this System 7. Note that the system obtained by setting $C=0$ in System 5
is equivalent to a special case of System 2 and that Theorems 2 and 3 apply to 
it. The parameters are related by $a=2(A+B)$, $b=2B$. The special cases in
Fig. 2 of \cite{levbaror} correspond to $a=1.8$, $b=1$ and $a=2.5$, $b=1.3$.
It follows from (\ref{z1evol3}) and (\ref{z1evol4}) that 
$\frac{d}{dt}(z_1+z_2)=-(z_1+z_2)+2$ and that $z_1+z_2$ tends to two as 
$t\to\infty$. Thus the dynamics is controlled by that on the invariant 
manifold $z_1=2-z_2$. With the choices which have been made it coincides
with the invariant manifold $S_1$ introduced earlier.

Now consider what happens when $C\ne 0$. Provided $C\le\min\{A,B\}$ the system
consisting of (\ref{z1evol}) and (\ref{z2evol}) is competitive. Thus each
solution $(z_1,z_2)$ must converge to a stationary solution as $t\to\infty$
\cite{hirsch}. It can be concluded that the corresponding solution 
$(x_1,x_2,x_3,x_4)$ tends to a stationary solution. Information about 
stationary points of the system can be obtained by stability considerations, 
as was mentioned following Theorem 3. This discussion is summed up in the 
following theorem:

\noindent
{\bf Theorem 4} Any solution $x_i$ $(i=1,2,3,4)$ of (\ref{abc1})-(\ref{abc4})
with parameter values satisfying $C\le\min\{A,B\}$ converges to a stationary 
solution as $t\to\infty$. For fixed values of $A$ and $B$ and $C$ sufficiently 
small the system has at most three stationary solutions and at most two stable 
stationary solutions. 

\vskip 10pt\noindent
The results which have been obtained are unfortunately not sufficient to
give a rigorous confirmation of the qualitative behaviour shown in Fig. 2a
and Fig. 2b of \cite{levbaror}. The condition $C<\min\{A,B\}$ is not satisfied
by the parameter values in Fig. 2a. It is satisfied in the case of Fig. 
2b of \cite{levbaror} but no information is obtained about the number of
solutions. 

\section{Dynamics in the absence of macrophages}\label{Tcellonly}

In this section it will be shown that all the dynamical behaviour which has 
been shown to occur in system (\ref{basic}) also occurs in a truncated model
where the influence of macrophages is ignored. The system has two unknowns 
$x_1$ and $x_2$ with the same interpretation as before. The effect of the 
macrophages is turned off by setting $a_{ij}=0$ when $i$ is one or two and
$j$ is three or four. This leads to the following closed system for $x_1$ 
and $x_2$:
\begin{eqnarray}
&&\frac{dx_1}{dt}=-d_1x_1+g(a_{11}x_1-a_{22}x_2)   \\
&&\frac{dx_2}{dt}=-d_2x_2+g(-a_{11}x_1+a_{22}x_2)
\end{eqnarray}
All the coeffients $d_1$, $d_2$, $a_{11}$ and $a_{22}$ are assumed positive,
as before. Restricting further to the case that $d_1=d_2=d$ and $\theta=0$
allows the analysis of the dynamics to be reduced to that of a scalar
equation as in the previous section. The scalar equation is
\begin{equation}
\frac{dx_1}{dt}=-dx_1+g((a_{11}+a_{22})x_1-d^{-1}a_{22}).
\end{equation}
This is essentially the equation (\ref{toy}) analysed before and so all the 
phenomena found previously occur here also.

\section{Further remarks}

In this paper it has been possible to give a rigorous analysis of the 
asymptotics of solutions of the system of Lev Bar-Or \cite{levbaror} for some 
values of the parameters. Unfortunately there are large ranges of the 
parameters for which no conclusions were obtained or for which those
conclusions are incomplete. The latter statement even includes the two 
cases for which numerical plots were included in \cite{levbaror}. The following 
questions have not been answered for the system (\ref{basic}) with general 
parameters.
\begin{itemize}
\item{Are there periodic solutions?}
\item{Are there homoclinic orbits?}
\item{Are there heteroclinic cycles?}
\item{Are there strange attractors?}
\end{itemize}
It should be emphasized that there is no evidence, analytical or numerical, 
that the answer to any of these questions is yes in the general case. We are 
left with a picture of a dynamical system where the long-time behaviour seems 
to be simple but it is quite unclear how to prove it except for a restricted 
set of values of the parameters.

It is consistent with everything which has been found here that the system
of four equations with eighteen parameters produces no qualitatively new 
phenomena in comparison with a reduced two-dimensional system with three
parameters. No effects were found which are specifically dependent on
the inclusion of the presentation of antigen by macrophages. The interactions
between Th1 and Th2 cells appear to be sufficient. It is seen that under
some circumstances the model with T cells and cytokines alone predicts
a situation of bistability where either a Th1 or Th2 dominated state can
be approached, depending on the initial data.

\end{document}